\documentclass[english,reprint,showkeys]{revtex4-1}
\usepackage[T1]{fontenc}
\usepackage[latin9]{inputenc}
\usepackage{amstext}
\usepackage{amssymb}
\usepackage{graphicx}

\makeatletter

\providecommand{\tabularnewline}{\\}

\makeatother

\usepackage{babel}
\begin{document}

\title{Short-term predictions of country-specific Covid-19 infection rates
based on power law scaling exponents}

\author{H.M. Singer}

\date{2020-03-25}

\affiliation{Institute for Complex Systems Research ICSR\break 8820 Wädenswil,
Switzerland}
\email{hsinger@icsr.ch}

\keywords{power-law scaling, data analysis, epidemic dynamics}
\begin{abstract}
The number of corona virus (COVID-19) infections grows worldwide.
In order to create short term predictions to prepare for the extent
of the global pandemic we analyze infection data from the top 25 affected
countries. It is shown that all country-specific infection rates follow
a power law growth behavior and the scaling exponents per country
are calculated. We find two different growth patterns: either steady
power law growth from the very beginning with moderate scaling exponents
of 3-5 or explosive power law growth with dramatic scaling exponents
of 8-11. In the case of the USA we even find an exponent of 16.59.
By means of data analysis we confirm that instituting strict measures
of lock-downs combined with a strong adherence by the population are
effective means to bring the growth rates down. While many countries
have instituted measures there are only three countries (Denmark,
Norway, and South Korea) so far where such lock-downs led to a significant
reduction of the growth rate. In the case of Denmark we calculate
the reduction of the scaling exponents to move from 6.82 to 1.47.
\end{abstract}
\maketitle

\section*{Introduction}

Since the identification of a novel novel corona virus (COVID-19)
in Wuhan, China in December 2019 the virus kept spreading rapidly
throughout the world resulting in a global pandemic. As of March 25th
10:30 CET 424,048 people in over 170 countries and regions have been
infected \cite{arcgis}. Due to its high infectivity no slowing down
of the spread is currently in sight. A medical cure or vaccine is
not available yet. This pandemic brings health care systems worldwide
to their limits. The mortality rate is calculated to be 2.5\% \cite{LAI2020105924}
(comparison: SARS 9.6\% \cite{SMITH20063113} and H1N1 influence 0.6\%\cite{VaillantEtAl2009}).
As the number of infections rises many governments around the world
have instituted drastic lock-downs and curfews and called for social
distancing and work from home to reduce the rate at which the virus
spreads. 

The severity of the infections with the COVID-19 is not evenly distributed
with respect to age. While children do not seem to be affected much
the risk of complications and negative progression seems to increase
with increasing age \textendash{} the elderly population (70+) being
affected most (5-11\% mortality rate) \cite{jcm9020523}. Preexisting
health conditions such as cardiovascular diseases, diabetes, respiratory
diseases or cancer \cite{DRIGGIN2020,jcm9020523} and smoking \cite{Vardavas2020}
seem to increase the risk of complications as well. 

In order for health care systems to be able to predict the short term
and longer term number of infected patients different models have
been used. Epidemiological analysis models assumes an exponential
growth and a fixed reproduction number $n$, i.e. the number of people
infected by a sick person. For $n>1$ and an incubation time $\tau$
the total number of infections is assumed to grow exponentially
\[
n^{\frac{t}{\tau}}=e^{at}
\]

This approach is the basis for a number of different predictive studies
such as \cite{ZHAO2020214,schmitt2020simplified}.

Other approaches to estimate and predict the number of infections
are for example based on 'mechanistic-statistical' approaches coupling
a SIR ODE model describing the unobserved epidemiological dynamics
combined with a probabilistic model describing the data acquisition
process and a statistical inference method \cite{roques2020mechanisticstatistical},
time series modeling \cite{deb2020time}, iterative maps \cite{botha2020simple},
or agent based simulations \cite{chang2020modelling} as well as logistic
maps\cite{Hermanowicz2020.02.04.20020461,ROOSA2020256}. 

The problem with exponential approaches is that it assumes that any
infected person will infect the same number $n$ more people not taking
into account the human tendency to live and work in relatively closed
groups. Therefore such models tend to overestimate the number of infected
people as time progresses, simply because there are no more new people
to infect in the individuals particular group. On the other hand if
the disease were to spread only at the periphery of a compact region
of infections then a parabolic growth $t^{2}$ would be expected \cite{br2020quadratic}.
It is thus to be assumed that after an initial short exponential growth
a slow down should be observed where the growth of newly infected
people increases less than exponentially but probably faster than
$t^{2}$ as humans are not generally limited to the confines of a
compact region. Given what is known about human interactions and structures
mostly showing small world phenomena and scale free networks \cite{Watts1998,Watts2004}
the assumption of a power law behavior is reasonable. Early analysis
in February 2020 \cite{Ziff2020.02.16.20023820} showed this indeed
to be true. After a short initial exponential growth period the number
of infections, as well as the number of recovered and deceased patients
follows a power law:
\[
n(t)=Bt^{\gamma}
\]

In the case of the infection spreading in China the exponent $\gamma$
was determined to be $2.27$ \cite{Ziff2020.02.16.20023820} and $2.48$
\cite{li2020scaling} (with more data). In an update to \cite{Ziff2020.02.16.20023820}
it was suggested to take the saturation and slow down in the rate
of new infections into account by tapering off the growth with an
exponential slow down as suggested in \cite{Vazquez_2006}
\[
n(t)=Ct^{\gamma_{cutoff}}e^{-\frac{t}{t_{0}}}
\]

with $\gamma_{cutoff}=3.09$ and $t_{0}=8.90$ days. As a second parameter
$t_{0}$ is introduced the original scaling exponent $\gamma$ changes
slightly to $\gamma_{cutoff}>\gamma$ as a result of the fitting procedure.

\section*{Method}

All data related to COVID-19 was downloaded from the publicly available
JHU-CSSE (2020) data source provided continuously by the Johns Hopkins
University Center for Systems Science and Engineering (JHUCSSE) \cite{csse_github}.
Aggregate data calculation for countries with different regions or
states were performed after collecting by a simple summation over
all associated regions or provinces (for the countries USA, China,
and Canada). The data for Australia was omitted because the aggregated
data in \cite{csse_github} did not correspond to the value given
in \cite{arcgis}.

We have selected the data sets with the 25 most infections as presented
in \cite{arcgis} on 2020-03-24 11:00 CET. The countries and their
infection counts are given in Table I.

\begin{table}
\caption{Countries with the most COVID-19 infections (2020-03-24)}
\begin{tabular}{|c|c|}
\hline 
Country & \#infections (2020-03-24)\tabularnewline
\hline 
\hline 
China & 81588\tabularnewline
\hline 
Italy & 63927\tabularnewline
\hline 
USA  & 46481\tabularnewline
\hline 
Spain  & 35212\tabularnewline
\hline 
Germany  & 30081\tabularnewline
\hline 
Iran  & 24811\tabularnewline
\hline 
France  & 20149\tabularnewline
\hline 
South Korea  & 9037\tabularnewline
\hline 
Switzerland  & 8795 \tabularnewline
\hline 
United Kingdom  & 6733\tabularnewline
\hline 
Netherlands  & 4767 \tabularnewline
\hline 
Austria  & 4742\tabularnewline
\hline 
Belgium  & 4269 \tabularnewline
\hline 
Norway  & 2647\tabularnewline
\hline 
Canada & 2088\tabularnewline
\hline 
Portugal  & 2060\tabularnewline
\hline 
Sweden  & 2059\tabularnewline
\hline 
Brazil  & 1960\tabularnewline
\hline 
Denmark  & 1703\tabularnewline
\hline 
Israel  & 1656\tabularnewline
\hline 
Malaysia  & 1624 \tabularnewline
\hline 
Turkey  & 1529 \tabularnewline
\hline 
Japan  & 1140\tabularnewline
\hline 
India  & 511\tabularnewline
\hline 
Russia  & 444\tabularnewline
\hline 
\end{tabular}

\end{table}

We have plotted the data for each country in a log-log plot and determined
the range of power-law behavior. From there the scaling exponent for
each country was extracted. As an example the data for Germany is
shown in a log-log plot in Fig. 1. The range used to extract the scaling
exponent is denoted with two vertical lines. The best fit line (determined
by linear regression) is superimposed on the data points. In Fig.
2 the short term prediction of the total number of infections is superimposed
onto the data points for the the next 7 days.

\begin{figure}
\caption{Log-log plot of the total number of infections for Germany as a function
of time. The range of scale free behavior is denoted with two vertical
lines. The best fit line (linear regression) is superimposed on the
data points.}

\includegraphics[scale=0.25]{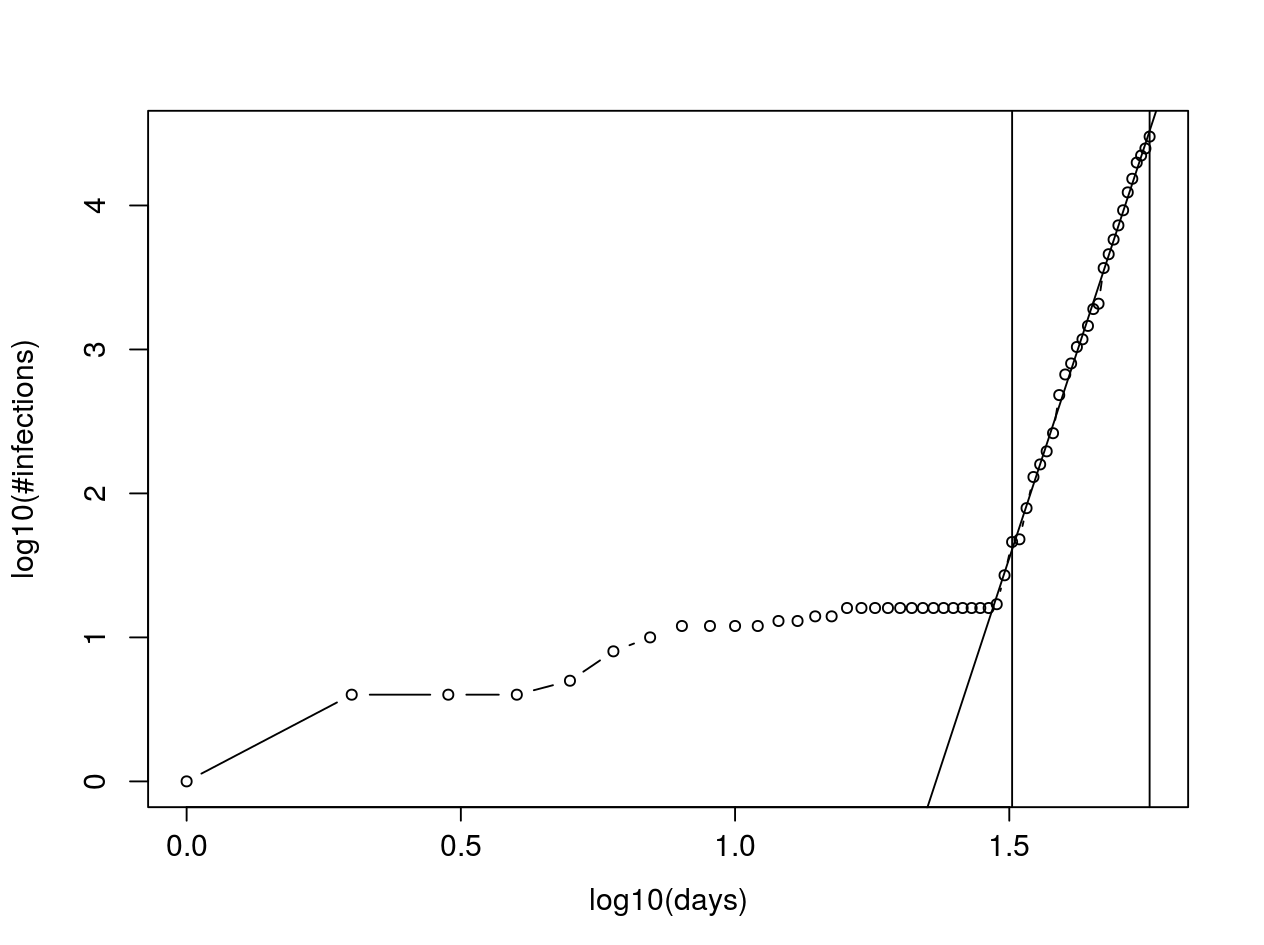}

\end{figure}

\begin{figure}
\caption{Short term prediction for the next 7 days of the total number of infections
for Germany superimposed on the available data points.}

\includegraphics[scale=0.25]{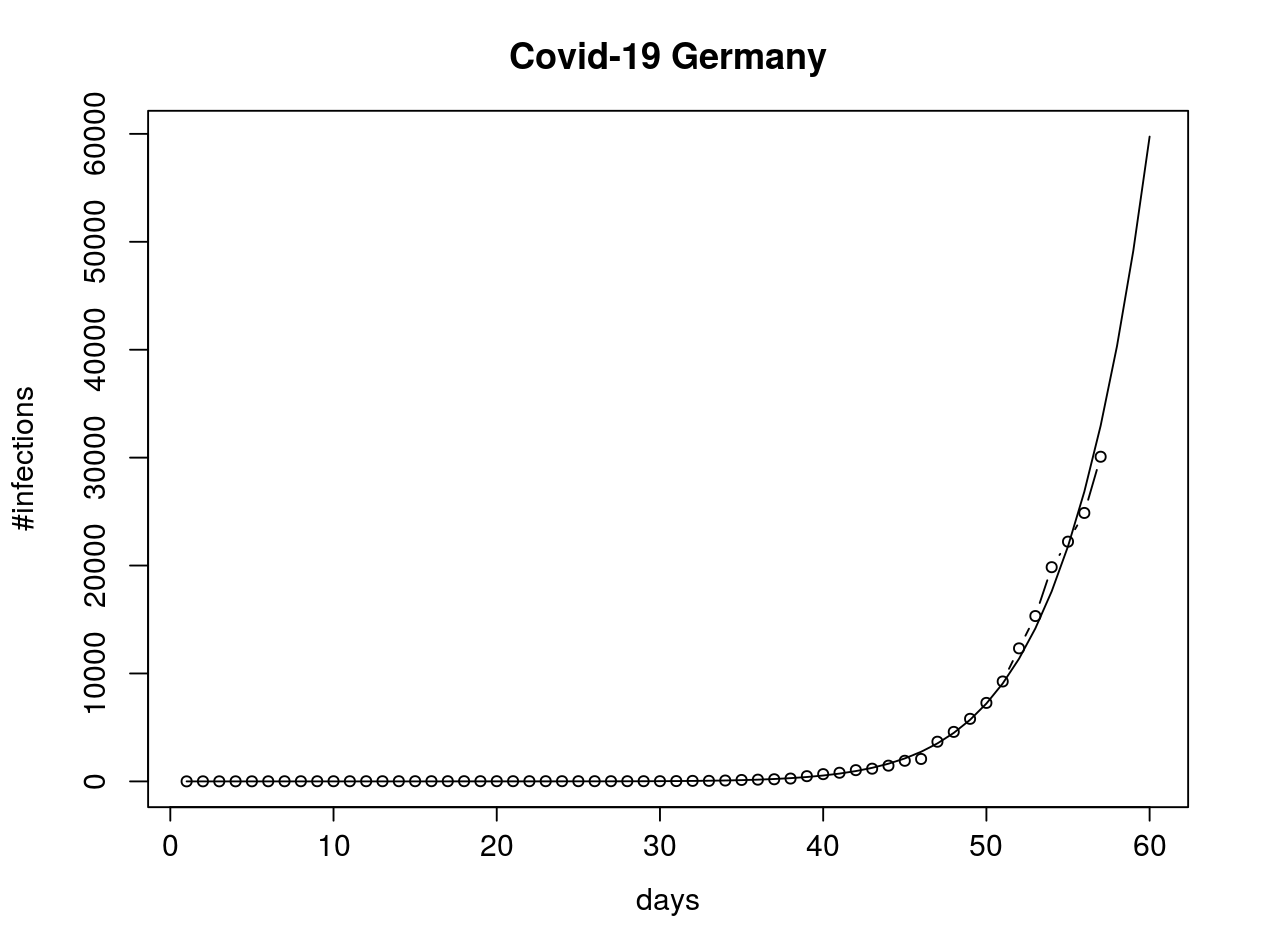}

\end{figure}

The scaling exponent for Germany was found to be $\gamma_{Germany}=11.59$. 

In order to cross-check the validity of the determined scaling exponent
we have recalculated the scaling exponent for China in the power law
range and found $\gamma_{China}=2.20$ in excellent agreement with
\cite{Ziff2020.02.16.20023820}. Fig. 3 shows the log-log plot for
China.

\begin{figure}
\includegraphics[scale=0.25]{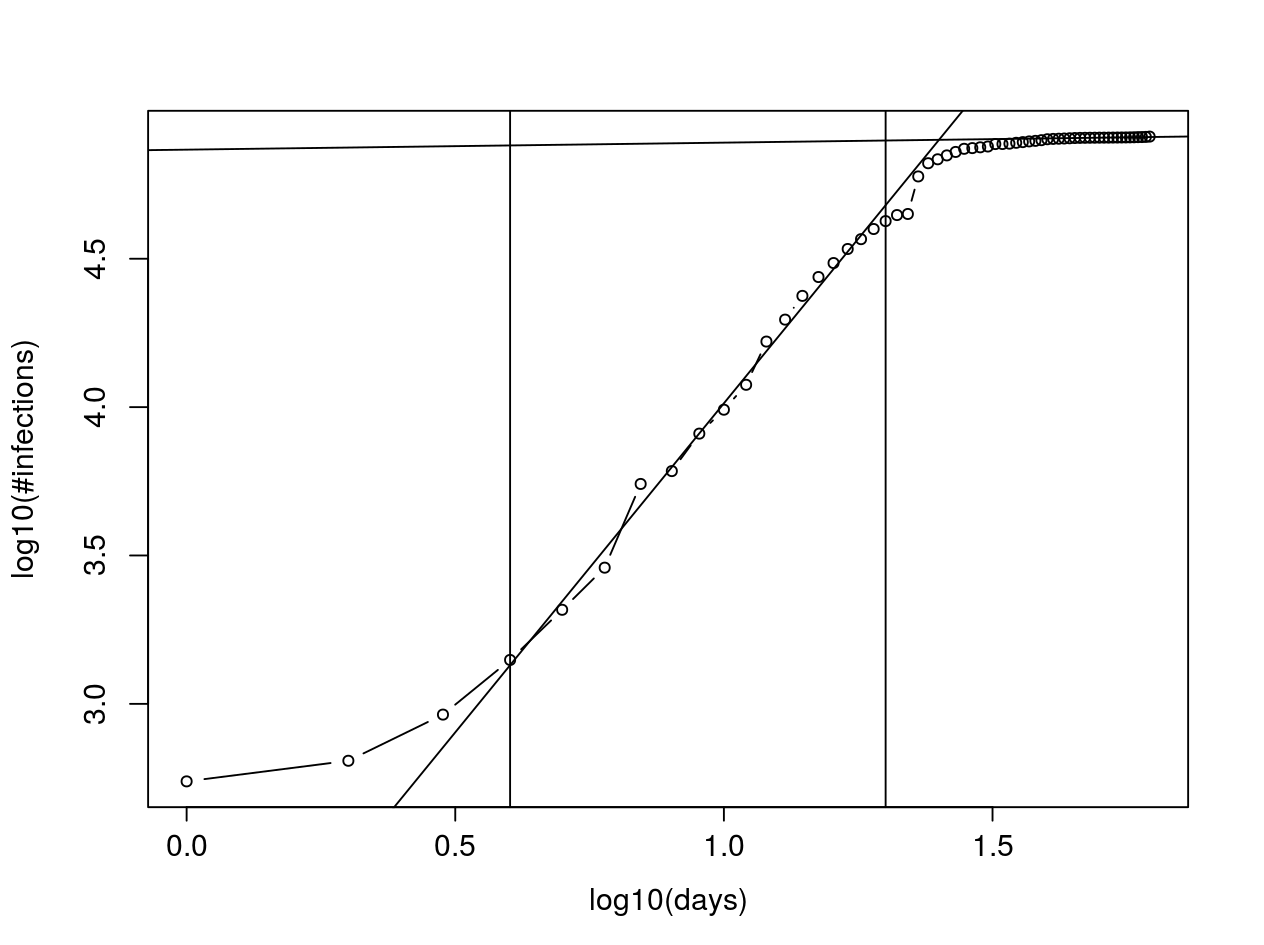}

\caption{Log-log plot of the total number of infections for China as a function
of time. The range of scale free behavior is denoted with two vertical
lines. The best fit line (linear regression) is superimposed on the
data points. A second fit line for the saturation range is superimposed
as well.}
\end{figure}

\begin{figure}
\caption{Available data for the total number of infections in China. Superimposed
are the scale free power law growth ($\gamma_{China,1}=2.20)$ and
the the saturation with $\gamma_{China,2}=0.024$.}

\includegraphics[scale=0.25]{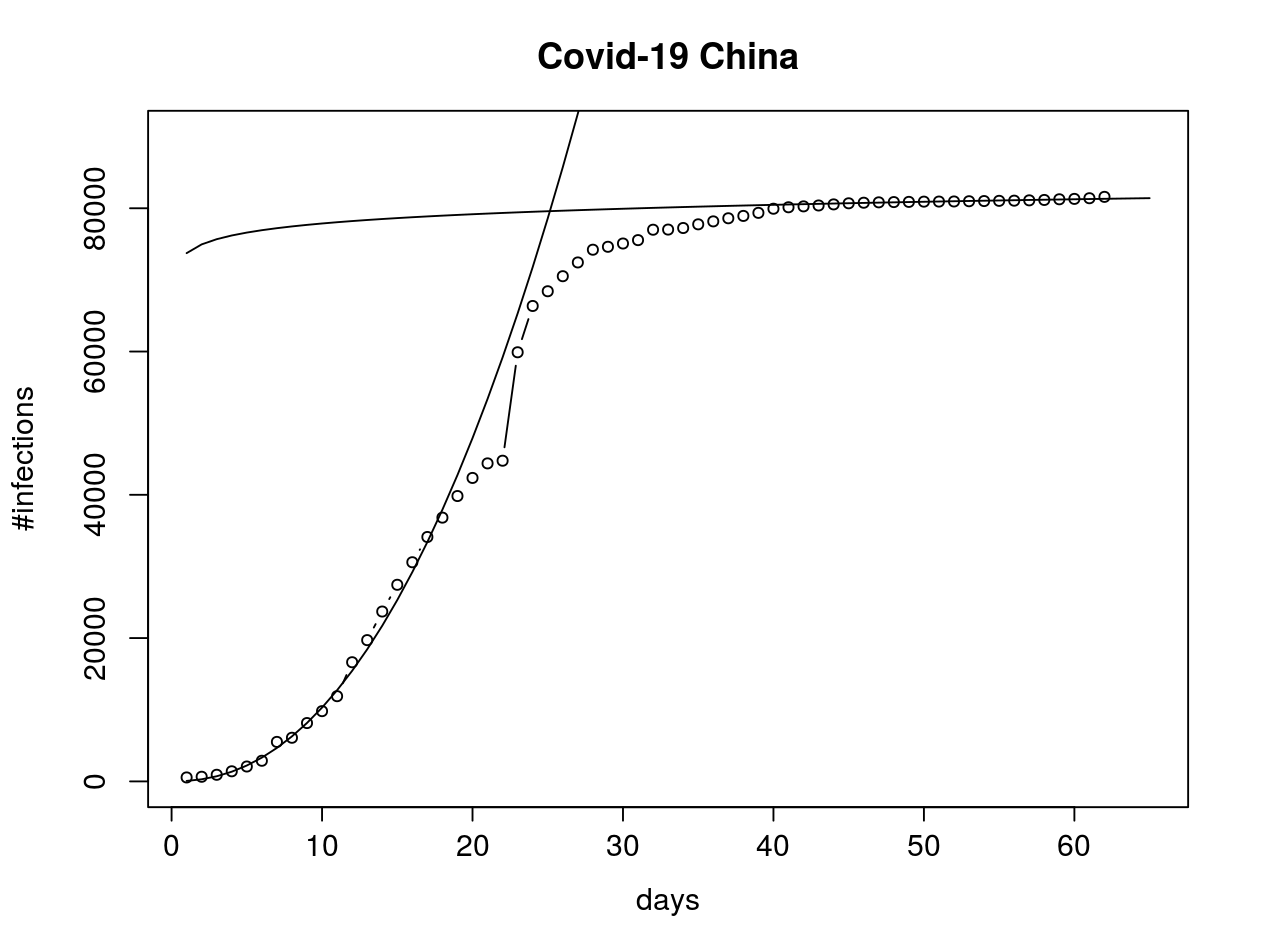}

\end{figure}

Since no country apart from China exhibits signs of a saturation we
have used the simple power law fit instead of the cutoff power law
fit \cite{Vazquez_2006} to determine all scaling exponents.

\section*{Results}

We have extracted the scaling exponents for all the countries in Table
I. As of 2020-03-24 there was only China which showed a saturation
of the total number of infections. All the other countries have not
reached yet such a level. We could distinguish three different stages:
\begin{enumerate}
\item \textbf{Growth stage:} the total number of infections is in the power
law range and a single scaling exponent could be extracted. Currently
most countries fall into this category.
\item \textbf{Slow down stage:} due to government imposed strict measures
to slow down the spread of the virus the initial power law range split
into a second one with a smaller scaling exponent. The total number
is still increasing as a power law but more slowly than before. We
have found only three countries where this was the case: Denmark,
Norway, and South Korea.
\item \textbf{Saturation stage:} the power law regime has ended and the
total number of infections slows down to small numbers close to 0.
This was only the case for China (see Fig. 3 and Fig. 4).
\end{enumerate}
To demonstrate the slow down stage we present the data from Denmark
in Fig. 5 as a log-log plot and Fig. 6 as short term predictions for
the total number of infections. We find $\gamma_{Denmark,1}=6.82$
pre-intervention and $\gamma_{Denmark,2}=1.47$ post-intervention.
It is, however, important to note that the data so far clearly does
not show a saturation \textendash{} merely a slow down into a different
power law regime with a different scaling exponent.

\begin{figure}
\includegraphics[scale=0.25]{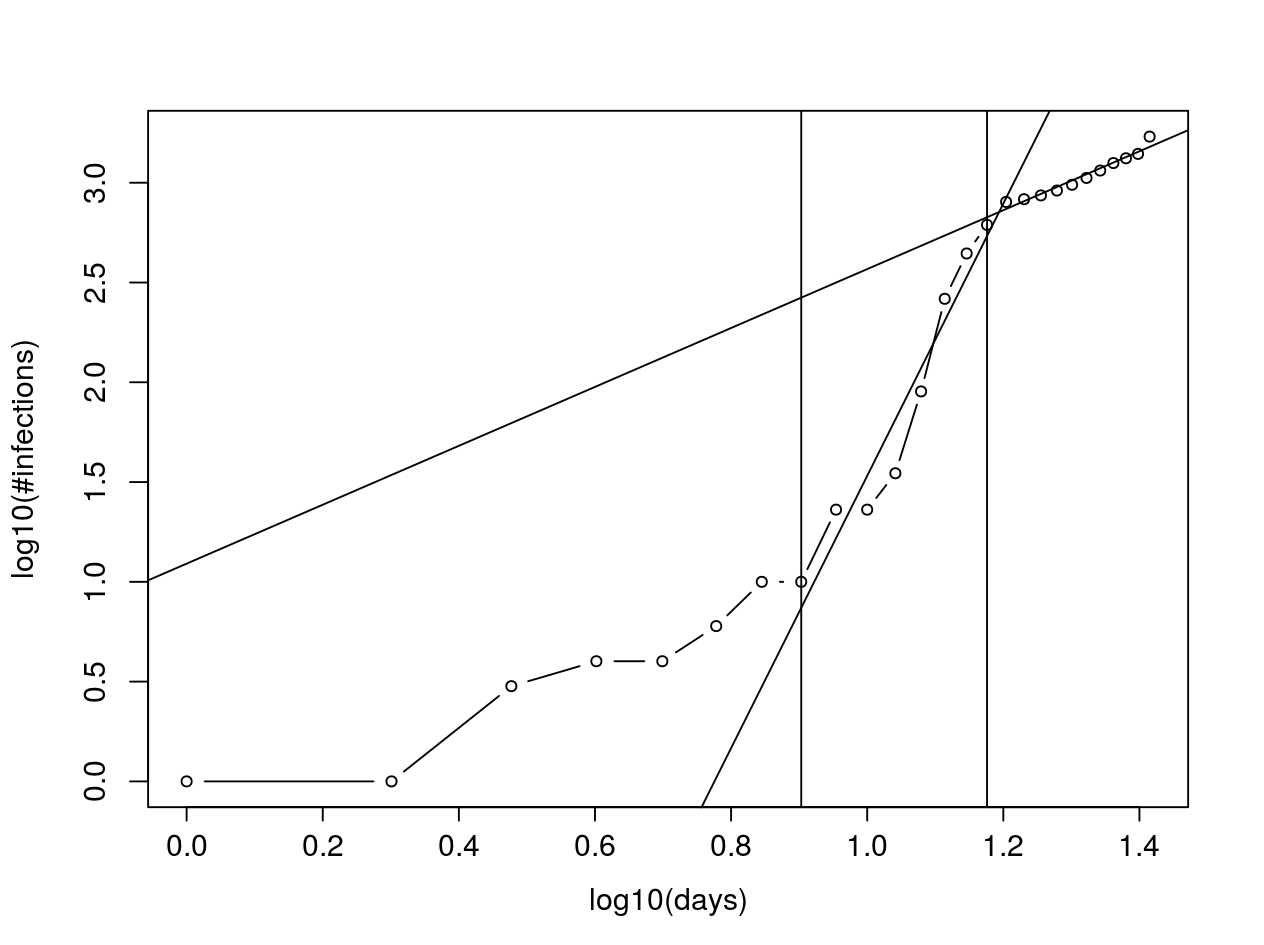}

\caption{Log-log plot of the total number of infections for Denmark as a function
of time. The two ranges of scale free behaviors are shown between
two vertical lines. The best fit lines (linear regressions) are superimposed
on the data points. We find $\gamma_{Denmark,1}=6.82$ pre-intervention
and $\gamma_{Denmark,2}=1.47$ post-intervention.}

\end{figure}

\begin{figure}
\includegraphics[scale=0.25]{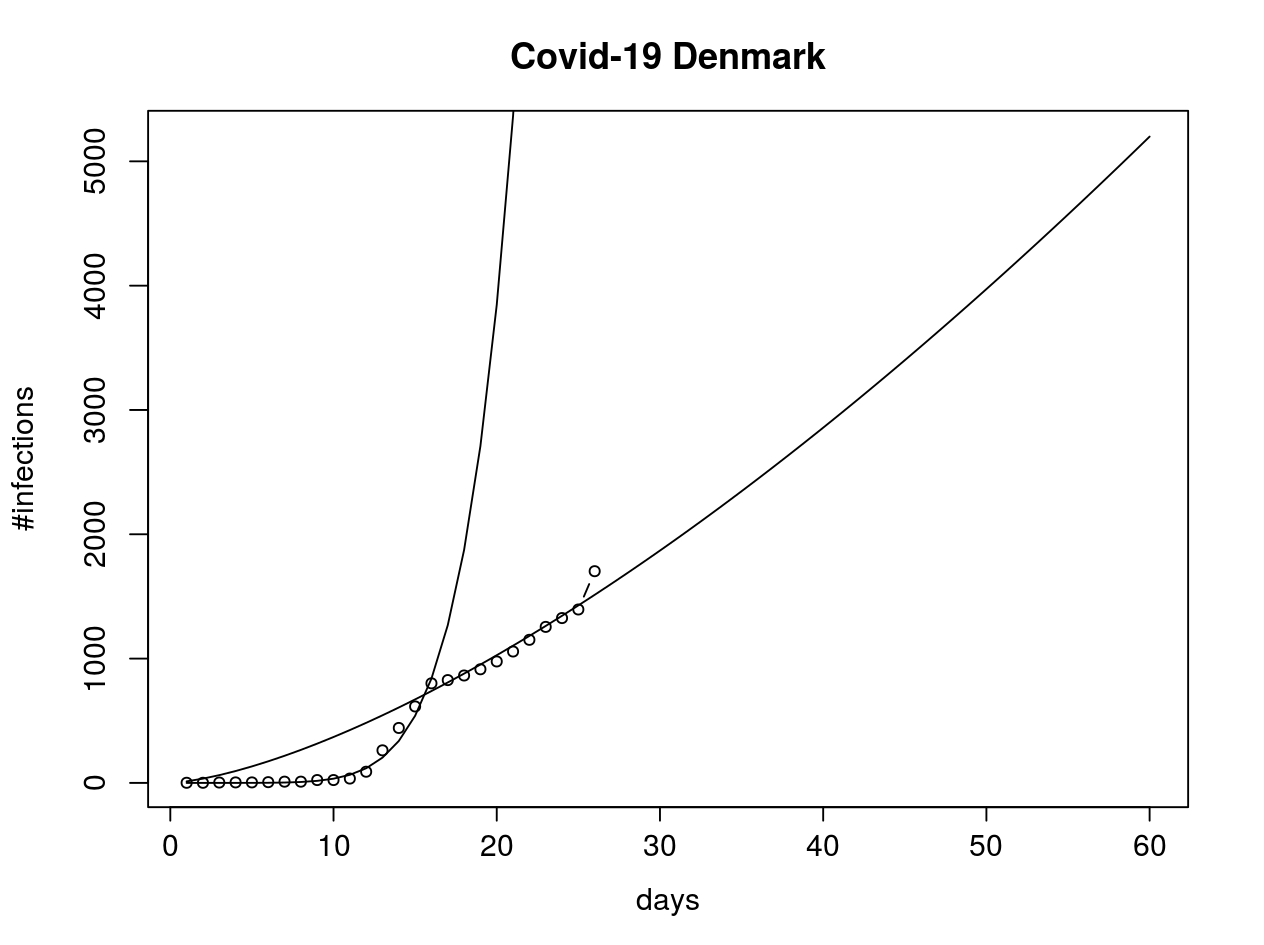}

\caption{Short term prediction for the next 30 days of the total number of
infections for Denmark. Superimposed are the scale free power law
growth $\gamma_{Denmark,1}=6.82$ and the post-intervention slow down
stage with $\gamma_{Denmark,2}=1.47$.}

\end{figure}

We have plotted all country data in a log-log plot starting from the
respective day of the first infection in Fig. 7. If more than one
person was recorded to be infected on the first day we have subtracted
the difference to 1 from the data set to plot all the data from the
same origin. Since we are only interested in the power law exponent
in the range of large total numbers of infections this subtraction
did not affect the outcome of the plot and the result. Since, however,
the number of recorded infections on the first day in China was 548
the data points from China do not follow the other countries' general
development.

As can be seen in Fig. 7 the growth rate of the total number of infections
can be roughly divided into two distinct growth patterns:
\begin{enumerate}
\item steady power law growth from the beginning in the blue (online) regime
where we have found a scaling exponent of $\gamma_{steady}\text{\ensuremath{\approx}}4.8\,(2-6)$.
\item explosive power law growth in the red (online) regime: after a long
incubation time with hardly any infections the total number of infections
increases rapidly with a scaling exponent of $\gamma_{explosive}\text{\ensuremath{\approx}}10\,(8-16)$.
\end{enumerate}
\begin{figure}
\includegraphics[scale=0.25]{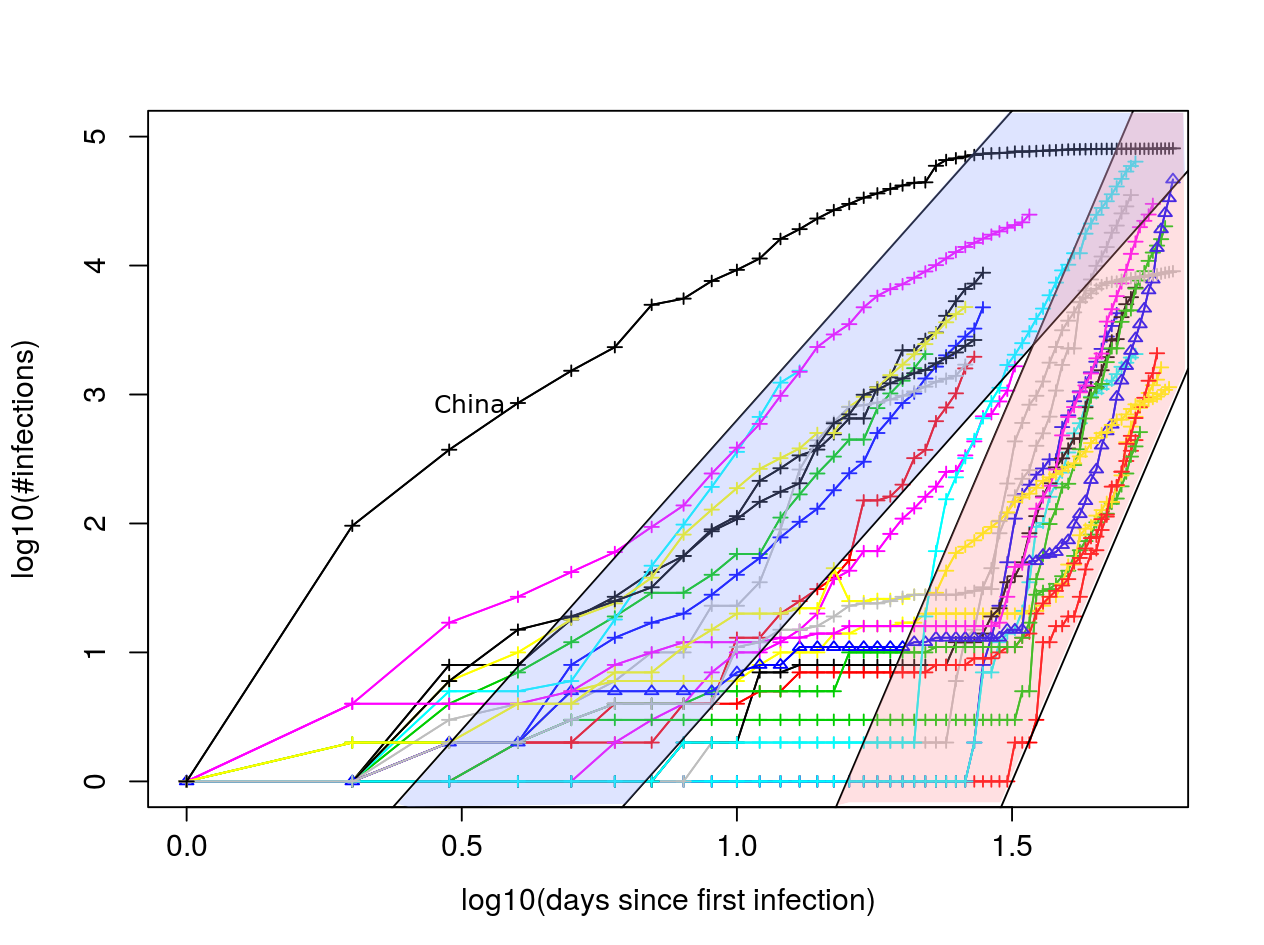}

\caption{All 25 country data superimposed in a log-log plot starting from the
day of the first infection. Since the first recorded data from China
is 548 infections the China data does not show the same increase as
the other curves that started with 1. Two distinct growth patterns
can be distinguished: blue range (online) steady power law growth
from the very beginning with a $\gamma_{steady}\text{\ensuremath{\approx}}4.8\,(2-6)$
and red range (online) with a long incubation time and an explosive
power law growth with a $\gamma_{explosive}\text{\ensuremath{\approx}}10\,(8-11)$.}

\end{figure}

The exact calculations of the power law scaling exponents are given
in Table II. The scaling exponents are displayed graphically in Fig.
8.

\begin{table}
\caption{Scaling exponents $\gamma$ per country. Countries with a slow down
stage or a saturation stage after government interventions and lock-downs
have a second scaling exponent $\gamma_{2}$ attached. The countries
are sorted according to increasing scaling exponent $\gamma$ from
steady growth to explosive growth. }
\begin{tabular}{|c|c|c|}
\hline 
Country & $\gamma$ (pre) & $\gamma_{2}$(post)\tabularnewline
\hline 
\hline 
China & 2.22 & 0.024 (saturation)\tabularnewline
\hline 
Iran & 2.26 & \tabularnewline
\hline 
Japan & 3.38 & \tabularnewline
\hline 
Netherlands & 3.63 & \tabularnewline
\hline 
Sweden & 4.50 & \tabularnewline
\hline 
Switzerland  & 4.72 & \tabularnewline
\hline 
Portugal & 4.89 & \tabularnewline
\hline 
Austria & 4.90 & \tabularnewline
\hline 
Norway & 5.25 & 2.08\tabularnewline
\hline 
Turkey & 5.91 & \tabularnewline
\hline 
Brazil & 6.39 & \tabularnewline
\hline 
Israel & 6.60 & \tabularnewline
\hline 
Italy & 6.80 & \tabularnewline
\hline 
Denmark & 6.82 & 1.48\tabularnewline
\hline 
Belgium & 8.88 & \tabularnewline
\hline 
Malaysia & 9.05 & \tabularnewline
\hline 
United Kingdom  & 9.62 & \tabularnewline
\hline 
India & 9.76 & \tabularnewline
\hline 
France & 10.14 & \tabularnewline
\hline 
Russia & 10.33 & \tabularnewline
\hline 
Spain & 10.34 & \tabularnewline
\hline 
South Korea & 10.88 & 0.83\tabularnewline
\hline 
Germany & 11.58 & \tabularnewline
\hline 
Canada & 11.67 & \tabularnewline
\hline 
USA & 16.59 & \tabularnewline
\hline 
\end{tabular}
\end{table}

\begin{figure}
\includegraphics[scale=0.35]{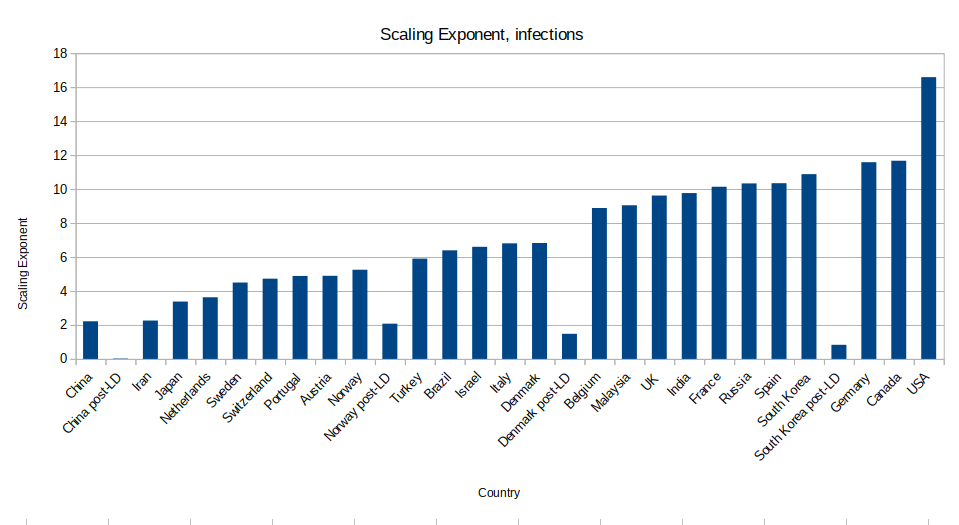}

\caption{Scaling exponents per country sorted in ascending order. Countries
with slow down stages or saturation stages have the post-intervention
lock-down (post-LD) scaling exponent $\gamma_{2}$ plotted next to
the inital $\gamma$.}

\end{figure}

\section*{Conclusions}

Our initial assumption was that cultural habits of proximity and cleanliness
such as the habit of washing hands as well as the geographical location
would influence the speed at which the total number of infection increases
but at least on a per country comparison such a distinction was not
discernible. Neither cultural habits nor social distance and personal
space \cite{social_dist} seem to have a noticeable influence. Also
the latitude and the according temperature differences do not seem
to play any role. 

A puzzling fact is that neighboring countries do not seem to exhibit
similar scaling exponents such as Belgium (8.88) to the Netherlands
(3.63) or France (10.14) to Switzerland (4.72) or Germany (11.58)
to Austria (4.9).

We have confirmed that the implementation of lock-downs had an impact
on the spread \textendash{} however, not the governmental sanctions
\textit{per se} but \textendash{} obviously \textendash{} the adherence
to it by the people. Notably there was hardly any noticeable impact
on the lock-down in Italy whereas the measures in Denmark and Norway
were very effective. The complete lock-down of Wuhan in China seemed
to have been so far the most effective measure, however, China is
two months ahead of all the other countries so no direct comparison
can be drawn at the moment. South Korea deserves a special mention
since the wide range testing of a large part of the population and
consequent isolation of confirmed cases seems to have prevented any
further growth (dropping from 10.88 to 0.83).

A puzzling data point is Iran which has the second lowest scaling
exponent (2.26) almost identical to China (2.20). As Iran to our knowledge
did not introduce any severe and strict measures and lock-downs we
are tempted to believe that the reported numbers do not correspond
to the actual numbers of infected people in that country. Perhaps
many people in Iran are not able to seek medical attention or prefer
not to?

It is worth noting that the currently highest spread is found in the
USA with a scaling exponent of 16.59. We do not know why the outbreak
in the US was so explosive.

We have found that the stricter the lock-down measures have been instituted
and the higher the adherence of the population to those measures the
faster the exponent dropped and the more noticeable pre- and post-measure
regimes were separated.

Countries where there were no strict measures (or not yet) and no
adherence to those measures if implemented or if they were introduced
very late in the process have scaling exponents of $\thicksim8-11$
(USA 16.59). Countries with earlier lock-down measures and stricter
adherence have exponents of $\thicksim3-5$.

In conclusion the data supports evidence that the spread is not exponential
but can be described as a power law which indicates that all investigated
countries and societies seem to be organized as small world networks
and show scale-free behavior in the total number of infections. Different
governmental measures and adherence of the population to those measures
leads to strikingly different growth rates and scaling exponents.
The power law approach fits the recorded data very well and allows
accurate short term predictions of the total number of infections
which allows health care systems to prepare and to effectively plan
necessary staff, infrastructure and in particular intensive care units
and to triage new patients more efficiently.
\begin{acknowledgments}
The author reports no funding related to this research and has no
conflicting financial interests.
\end{acknowledgments}

\bibliographystyle{unsrtnat}
\bibliography{covid01}

\end{document}